\documentclass[11pt]{article}           %[9pt,bezier]
\usepackage{amsmath}
\usepackage{amssymb}
\newtheorem{Theorem}{Theorem}
\newtheorem{Remark}{Remark}

\usepackage{graphics}

\date{}

\title{BINARY NONLINEARIZATION OF THE SUPER AKNS SYSTEM UNDER AN IMPLICIT SYMMETRY CONSTRAINT}
\author{
 JING YU$^{a}$, JINGWEI HAN$^{a}$,
JINGSONG HE$^{b,c}$\footnote{Corresponding author, E-mail
address:jshe@ustc.edu.cn, hejingsong@nbu.edu.cn}
\vspace{4mm}\\
$^{a}$School of Science, Hangzhou Dianzi University, Hangzhou,
Zhejiang, 310018,  China\\
$^{b}$Department of Mathematics, University of Science and
Technology of China, Hefei, Anhui,\\ 230026,  China\\
$^{c}$Department of Mathematics, Ningbo University, Ningbo,
Zhejiang, 315211,
 China}
\begin{document}
 \maketitle

 \begin{abstract}
For the super AKNS system, an implicit symmetry constraint between
the potentials and the eigenfunctions is proposed. After introducing
some new variables to explicitly express potentials, the super AKNS
system is decomposed into two compatible finite-dimensional super
systems (x-part and $t_n$-part). Furthermore, we show that the
obtained  super systems are  integrable super Hamiltonian systems in
supersymmetry manifold $\mathbb{R}^{4N+2|2N+2}$.

%\vspace{1.5cm}


\end{abstract}

\section{Introduction}

In 1988, Prof. Cao proposed the mono-nonlinearization method of Lax
pairs for the classical integrable (1+1)-dimensional system
\cite{cao88}. The key of mono-nonlinearization is to find the
constraint between the potentials and the eigenfunctions of the Lax
system. After choosing some distinct spectral parameters and
considering the constraint, the Lax system is decomposed into
finite-dimensional systems whose variables can be separated, and
furthermore, the obtained finite-dimensional systems are completely
integrable Hamiltonian systems in the Liouville sense. Several years
later, the method was extended to classical integrable
(2+1)-dimensional systems \cite{Konopelchenko,cheng91,cheng92}.
Thereafter, the method was continued to generalize in the following
two aspects. One was the binary nonlinearization method of the Lax
pairs and its adjoint Lax pairs for the classical integrable
systems, which was firstly proposed by Prof. Ma in Ref. \cite{ma94}.
And the other was the higher-order constraints (i. e. implicit
constraints), which were widely studied in Refs.
\cite{zeng95,ma00,yu08}. In a word, the method of nonlinearization
was extensively studied by many researchers in the past twenty
years. Followed from this, many finite-dimensional integrable
Hamiltonian systems were obtained.

In recent years, several integrable super  systems
\cite{Gurses85,Li86,Shaw98} and integrable supersymmetry  systems
\cite{popowicz,aratyn99,ghost03} have aroused strong interests in
many mathematicians and physicists. Such as Darboux transformation
\cite{liu97,liu00}, Hamiltonian structures \cite{li90,oevel91,tu99},
and so on. In very recent years, nonlinearization of the super AKNS
system has been studied  in Ref. \cite{Heyu}, where we considered an
explicit symmetry constraint of the super AKNS system, and we proved
that under the explicit constraint, the super AKNS system was
completely integrable super Hamiltonian system in the Liouville
sense. Inspired by this and implicit constraint of the classical
integrable system, a natural question is appeared whether an
implicit symmetry constraint is available for the super AKNS system.
In the present paper, we shall solve this problem.

The paper is organized as follows. In the next section,  we propose
an implicit symmetry constraint between the potentials and the
eigenfunctions of the super AKNS system. Then in section 3, under
the constraint, the super AKNS system is decomposed into two
compatible finite-dimensional super  systems. And furthermore, we
show that the obtained  finite-dimensional super systems are
completely integrable in the Liouville sense. Finally,  some
conclusions and discussions are listed in section 4.

\section{An Implicit Symmetry Constraint Of The Super AKNS
Hierarchy}

In Ref. \cite{Heyu}, we have considered the binary nonlinearization
of the super AKNS system under an explicit symmetry constraint, and
obtained the finite-dimensional integrable super Hamiltonian system.
Here we will propose an implicit symmetry constraint of the super
AKNS system. Therefore, this paper can be regarded as a continuation
 of Ref. \cite{Heyu}. For simplicity, we omit the
detailed derivation of the super AKNS hierarchy, which can be
referred to Ref. \cite{Heyu}. In what follows,  we shall propose an
implicit symmetry constraint between the potential  and the
eigenfunctions. To this aim, we consider the super AKNS spectral
problem
\begin{equation}\label{1}
\phi_x=M\phi, \quad  M=\left(\begin{array}{ccc}
-\lambda&q&\alpha\\
r&\lambda&\beta\\
\beta&-\alpha&0\\
 \end{array}\right),\quad
 \phi= \left(\begin{array}{ccc}
\phi_1\\
\phi_2\\
\phi_3
 \end{array}\right),\end{equation}and
its adjoint spectral problem \begin{equation}\label{14}
\psi_x=-M^{St}\psi=\left(\begin{array}{ccc}\lambda&-r&\beta\\-q&-\lambda&-\alpha\\-\alpha&-\beta&0\end{array}\right)\psi,
\quad
\psi=\left(\begin{array}{ccc}\psi_1\\\psi_2\\\psi_3\end{array}\right),\end{equation}
where St means supertranspose \cite{cartier}.

By a similar way of the counterpart in the classical system
\cite{ma94, ma95}, it is not difficult to get the variational
derivative of the parameter  with respect to the potential
\begin{eqnarray}\label{16}
\frac{\delta\lambda}{\delta U_0}=\left(\begin{array}{c}
\frac{\delta\lambda}{\delta r}\\
\frac{\delta\lambda}{\delta q}\\
\frac{\delta\lambda}{\delta \beta}\\
\frac{\delta\lambda}{\delta \alpha}\end{array}\right)=
\left(\begin{array}{c}\psi_2\phi_1\\
\psi_1\phi_2\\
\psi_2\phi_3-\psi_3\phi_1\\
\psi_1\phi_3+\psi_3\phi_2\end{array}\right),
\end{eqnarray}
where $U_0=(r, q, \beta, \alpha)^{T}$. Imposing the zero boundary
conditions
 $\lim_{|x|\rightarrow\infty}\phi_i=\lim_{|x|\rightarrow\infty}\psi_i=0 (i=1, 2, 3)$, we can verify a simple characteristic property
\begin{equation}\label{14*}
L_1\frac{\delta\lambda}{\delta
U_0}=\lambda\frac{\delta\lambda}{\delta U_0},\end{equation} where
\begin{equation}\label{L1}L_1=\left(\begin{array}{cccc}
-\frac{1}{2}\partial_x+q\partial_x^{-1}r&-q\partial_x^{-1}q&
\frac{1}{2}\alpha+\frac{1}{2}q\partial_x^{-1}\beta&-\frac{1}{2}q\partial_x^{-1}\alpha\\
r\partial_x^{-1}r&\frac{1}{2}\partial_x-r\partial_x^{-1}q&\frac{1}{2}r\partial_x^{-1}\beta
&-\frac{1}{2}\beta-\frac{1}{2}r\partial_x^{-1}\alpha\\
2\beta-2\alpha\partial_x^{-1}r&2\alpha\partial_x^{-1}q&-\partial_x-\alpha\partial_x^{-1}\beta&
-q+\alpha\partial_x^{-1}\alpha\\
2\beta\partial_x^{-1}r&2\alpha-2\beta\partial_x^{-1}q&r+\beta\partial_x^{-1}\beta&
\partial_x-\beta\partial_x^{-1}\alpha\end{array}\right),\end{equation}
with $\partial_x=\frac{d}{dx},
\partial_x\partial_x^{-1}=\partial_x^{-1}\partial_x=1.$

 Choosing N distinct spectral parameters
$\lambda_1,\cdots,\lambda_N$, the super AKNS spectral problem
(\ref{1}) and the adjoint spectral problem (\ref{14}) become the
following finite-dimensional super systems
\begin{equation}\label{16*}\left\{\begin{array}{l}
\phi_{1j,
x}=-\lambda_j\phi_{1j}+q\phi_{2j}+\alpha\phi_{3j},\quad 1\leq j\leq N,\\
\phi_{2j,
x}=r\phi_{1j}+\lambda_j\phi_{2j}+\beta\phi_{3j},\quad 1\leq j\leq N,\\
\phi_{3j, x}=\beta\phi_{1j}-\alpha\phi_{2j},
\quad 1\leq j\leq N,\\
\psi_{1j, x}=\lambda_j\psi_{1j}-r\psi_{2j}+\beta\psi_{3j},
\quad 1\leq j\leq N,\\
\psi_{2j,
x}=-q\psi_{1j}-\lambda_j\psi_{2j}-\alpha\psi_{3j},\quad 1\leq j\leq N,\\
\psi_{3j, x}=-\alpha\psi_{1j}-\beta\psi_{2j},\quad 1\leq j\leq
N.\end{array}\right.\end{equation} In what follows, let us consider
the traditional symmetry constraints
\begin{equation}\label{22}
\left(\begin{array}{cccc}b_{k+1}\\c_{k+1}\\-2\rho_{k+1}\\2\delta_{k+1}\end{array}\right)
=\sum\limits_{j=1}^{N}\gamma_j\left(\begin{array}{cccc}
\frac{\delta\lambda_j}{\delta r}\\
\frac{\delta\lambda_j}{\delta q}\\
\frac{\delta\lambda_j}{\delta\beta}\\
\frac{\delta\lambda_j}{\delta\alpha}\end{array}\right),\end{equation}
where $\gamma_j (1\leq j\leq N)$ are usual constants and $k\geq0$.
In Ref. \cite{Heyu}, we have chosen $k=0$ and $\gamma_j=1 (1\leq
j\leq N)$ in the above constraint. Thus we obtained an explicit
symmetry constraint (i. e. the potentials can be expressed by the
eigenfunctions explicitly). While in this paper, we will extend our
previously work and choose $k=1$ and $\gamma_j=-\frac{1}{2} (1\leq
j\leq N)$ in Eq. (\ref{22}). That is to say, we obtain the following
implicit symmetry constraint
\begin{equation}\label{15}\left\{\begin{array}{l}
q_x=<\Psi_2, \Phi_1>,\\
r_x=-<\Psi_1, \Phi_2>,\\
\alpha_x=-\frac{1}{4}(<\Psi_2, \Phi_3>-<\Psi_3,
\Phi_1>),\\
\beta_x=-\frac{1}{4}(<\Psi_1, \Phi_3>+<\Psi_3,
\Phi_2>),\end{array}\right.\end{equation} where $\Phi_i=(\phi_{i1},
\cdots, \phi_{iN})^{T}, \Psi_i=(\psi_{i1}, \cdots, \psi_{iN})^{T},
i=1, 2, 3$, and $<.,.>$ denotes the standard inner product in
Euclidean space $R^{N}$. Obviously, the constraint (\ref{15}) is an
implicit constraint. That is to say, the potentials of the
finite-dimensional super systems (\ref{16*}) can not be expressed by
the eigenfunctions explicitly, which is different from the
constraint in Ref. \cite{Heyu}.  In order to consider
nonlinearization of the super AKNS system under the implicit
symmetry constraint (\ref{15}), we should take some measures.

\section{Nonlinearization of the super AKNS system under an implicit symmetry constraint}

Now we are in a position to discuss the nonlinearization of the
super AKNS system under the implicit symmetry constraint (\ref{15}).
To this aim, we firstly introduce the following new variables
\begin{equation}\label{new viables} \phi_{N+1}=q,\quad
\phi_{N+2}=2\alpha, \quad \psi_{N+1}=r,\quad
\psi_{N+2}=-2\beta.\end{equation} Considering the new variables
(\ref{new viables}) and substituting the constraint (\ref{15}) into
system (\ref{16*}), we obtain the following finite-dimensional super
system
\begin{equation}\label{16}\left\{\begin{array}{l}
\phi_{1j,
x}=-\lambda_j\phi_{1j}+\phi_{N+1}\phi_{2j}+\frac{1}{2}\phi_{N+2}\phi_{3j},\quad 1\leq j\leq N,\\
\phi_{2j,
x}=\psi_{N+1}\phi_{1j}+\lambda_j\phi_{2j}-\frac{1}{2}\psi_{N+2}\phi_{3j},\quad 1\leq j\leq N,\\
\phi_{3j,
x}=-\frac{1}{2}\psi_{N+2}\phi_{1j}-\frac{1}{2}\phi_{N+2}\phi_{2j},
\quad 1\leq j\leq N,\\
\phi_{N+1, x}=<\Psi_2, \Phi_1>,\\
\phi_{N+2, x}=-\frac{1}{2}(<\Psi_2, \Phi_3>-<\Psi_3, \Phi_1>),\\
\psi_{1j,
x}=\lambda_j\psi_{1j}-\psi_{N+1}\psi_{2j}-\frac{1}{2}\psi_{N+2}\psi_{3j},
\quad 1\leq j\leq N,\\
\psi_{2j,
x}=-\phi_{N+1}\psi_{1j}-\lambda_j\psi_{2j}-\frac{1}{2}\phi_{N+2}\psi_{3j},\quad 1\leq j\leq N,\\
\psi_{3j, x}=-\frac{1}{2}\phi_{N+2}
\psi_{1j}+\frac{1}{2}\psi_{N+2}\psi_{2j},\quad 1\leq j\leq
N,\\
\psi_{N+1, x}=-<\Psi_1, \Phi_2>,\\
\psi_{N+2, x}=\frac{1}{2}(<\Psi_1, \Phi_3>+<\Psi_3,
\Phi_2>).\end{array}\right.\end{equation} Obviously, Eq. (\ref{16})
can be written by the following super Hamiltonian form:
\begin{equation}\left\{\begin{array}{l}
\Phi_{1, x}=\frac{\partial H_1}{\partial\Psi_1},\quad \Phi_{2,
x}=\frac{\partial H_1}{\partial\Psi_2},\quad\Phi_{3,
x}=\frac{\partial H_1}{\partial\Psi_3},\quad \phi_{N+1,
x}=\frac{\partial H_1}{\partial\psi_{N+1}},\quad \phi_{N+2,
x}=\frac{\partial H_1}{\partial\psi_{N+2}},\\
\Psi_{1, x}=-\frac{\partial H_1}{\partial\Phi_1},\quad \Psi_{2,
x}=-\frac{\partial H_1}{\partial\Phi_2},\quad\Psi_{3,
x}=\frac{\partial H_1}{\partial\Phi_3},\quad \psi_{N+1,
x}=-\frac{\partial H_1}{\partial\phi_{N+1}},\quad \psi_{N+2,
x}=\frac{\partial
H_1}{\partial\phi_{N+2}},\end{array}\right.\end{equation}
 with the super
Hamiltonian is given by
\begin{eqnarray*}
H_1&=&-<\Lambda\Psi_1, \Phi_1>+<\Lambda\Psi_2,
\Phi_2>+\phi_{N+1}<\Psi_1, \Phi_2>+\psi_{N+1}<\Psi_2,
\Phi_1>\\&&+\frac{1}{2}\phi_{N+2}(<\Psi_1, \Phi_3>+<\Psi_3,
\Phi_2>)-\frac{1}{2}\psi_{N+2}(<\Psi_2, \Phi_3>-<\Psi_3,
\Phi_1>).\end{eqnarray*} That is to say, the nonlinearized
finite-dimensional super system (\ref{16}) is a super Hamiltonian
system.

In what follows, let us consider the temporal part of the super AKNS
hierarchy
\begin{equation}\label{5}
\phi_{t_n}=N^{(n)}\phi=(\lambda^nN)_{+}\phi,
\end{equation}
with
$$(\lambda^nN)_{+}=\sum\limits_{j=0}^{n}\left(\begin{array}{ccc}
a_j&b_j&\rho_j\\
c_j&-a_j&\delta_j\\
\delta_j&-\rho_j&0\end{array}\right)\lambda^{n-j},$$ where the
symbol "+" denotes taking the nonnegative power of $\lambda$. When
considered N distinct spectral parameter $\lambda_1, \cdots,
\lambda_N$, the temporal part of the super AKNS system becomes the
following super system
\begin{equation}\label{34}
\left(\begin{array}{c}
\phi_{1j}\\\phi_{2j}\\\phi_{3j}\end{array}\right)_{t_n}
=\left(\begin{array}{ccc}
\sum\limits_{i=0}^{n}a_i\lambda_j^{n-i}&\sum\limits_{i=0}^{n}b_i\lambda_j^{n-i}
&\sum\limits_{i=0}^{n}\rho_i\lambda_j^{n-i}\\\sum\limits_{i=0}^{n}c_i\lambda_j^{n-i}
&-\sum\limits_{i=0}^{n}a_i\lambda_j^{n-i}&\sum\limits_{i=0}^{n}\delta_i\lambda_j^{n-i}\\
\sum\limits_{i=0}^{n}\delta_i\lambda_j^{n-i}&-\sum\limits_{i=0}^{n}\rho_i\lambda_j^{n-i}&0
\end{array}\right)\left(\begin{array}{c}\phi_{1j}\\\phi_{2j}\\\phi_{3j}\end{array}\right)
,\quad 1\leq j\leq N,\end{equation} whose adjoint super system is
given by\begin{equation}\label{34*} \left(\begin{array}{c}
\psi_{1j}\\\psi_{2j}\\\psi_{3j}\end{array}\right)_{t_n}
=\left(\begin{array}{ccc}
-\sum\limits_{i=0}^{n}a_i\lambda_j^{n-i}&-\sum\limits_{i=0}^{n}c_i\lambda_j^{n-i}
&\sum\limits_{i=0}^{n}\delta_i\lambda_j^{n-i}\\-\sum\limits_{i=0}^{n}b_i\lambda_j^{n-i}
&\sum\limits_{i=0}^{n}a_i\lambda_j^{n-i}&-\sum\limits_{i=0}^{n}\rho_i\lambda_j^{n-i}\\
-\sum\limits_{i=0}^{n}\rho_i\lambda_j^{n-i}&-\sum\limits_{i=0}^{n}\delta_i\lambda_j^{n-i}&0
\end{array}\right)\left(\begin{array}{c}\psi_{1j}\\\psi_{2j}\\\psi_{3j}\end{array}\right)
,\quad 1\leq j\leq N.\end{equation} For $n=1$, systems (\ref{34})
and (\ref{34*}) are exactly the spatial systems (\ref{1}) and
(\ref{14}), respectively. In particular, as for $t_2$-part, the
nonlinearized super systems {(\ref{34}) and (\ref{34*}) become the
following system
\begin{equation}\label{21*}\left\{\begin{array}{l}
\phi_{1j,
t_2}=(-\lambda_j^{2}+\frac{1}{2}qr+\alpha\beta)\phi_{1j}+(q\lambda_j-\frac{1}{2}q_x)\phi_{2j}+(\alpha\lambda_j-\alpha_x)\phi_{3j},\quad1\leq j\leq N,\\
\phi_{2j, t_2}=(r\lambda_j+\frac{1}{2}r_x)\phi_{1j}+(\lambda_j^{2}-
\frac{1}{2}qr-\alpha\beta)\phi_{2j}+(\beta\lambda_j+\beta_x)\phi_{3j},\quad1\leq j\leq N,\\
\phi_{3j,
t_2}=(\beta\lambda_j+\beta_x)\phi_{1j}+(-\alpha\lambda_j+\alpha_x)\phi_{2j},\quad1\leq j\leq N,\\
\psi_{1j, t_2}=(\lambda_j^{2}-\frac{1}{2}qr-\alpha\beta)\psi_{1j}
-(r\lambda_j+\frac{1}{2}r_x)\psi_{2j}+(\beta\lambda_j++\beta_x)\psi_{3j},\quad1\leq j\leq N,\\
\psi_{2j, t_2}=(-q\lambda_j+\frac{1}{2}q_x)\psi_{1j}
+(-\lambda_j^{2}+\frac{1}{2}qr+\alpha\beta)\psi_{2j}+(-\alpha\lambda_j+\alpha_x)\psi_{3j},\quad1\leq j\leq N,\\
\psi_{3j,
t_2}=(-\alpha\lambda_j+\alpha_x)\psi_{1j}-(\beta\lambda_j+\beta_x)\psi_{2j},\quad1\leq
j\leq N.\end{array}\right.\end{equation} Considering the new
variables (\ref{new viables}) and the implicit constraint
(\ref{15}), the above finite-dimensional super system (\ref{21*})
becomes the following the nonlinearized super system
\begin{equation}\label{21}\left\{\begin{array}{l}
\phi_{1j,
t_2}=(-\lambda_j^{2}+\frac{1}{2}\phi_{N+1}\psi_{N+1}-\frac{1}{4}\phi_{N+2}\psi_{N+2})\phi_{1j}+(\phi_{N+1}\lambda_j-\frac{1}{2}<\Psi_2,
\Phi_1>)\phi_{2j}\\\qquad\quad+\frac{1}{4}(2\phi_{N+2}\lambda_j+<\Psi_2,
\Phi_3>-<\Psi_3, \Phi_1>)\phi_{3j},\\
\phi_{2j, t_2}=(\psi_{N+1}\lambda_j-\frac{1}{2}<\Psi_1,
\Phi_2>)\phi_{1j}+(\lambda_j^{2}-
\frac{1}{2}\phi_{N+1}\psi_{N+1}+\frac{1}{4}\phi_{N+2}\psi_{N+2})\phi_{2j}\\\qquad\quad
-\frac{1}{4}(2\psi_{N+2}\lambda_j+<\Psi_1, \Phi_3>+<\Psi_3,
\Phi_2>)\phi_{3j},\\
\phi_{3j, t_2}=-\frac{1}{4}(2\psi_{N+2}\lambda_j+<\Psi_1,
\Phi_3>+<\Psi_3,
\Phi_2>)\phi_{1j}-\frac{1}{4}(2\phi_{N+2}\lambda_j+<\Psi_2,
\Phi_3>-<\Psi_3, \Phi_1>)\phi_{2j},\\
\phi_{N+1, t_2}=\frac{1}{2}\phi_{N+1}(<\Psi_1, \Phi_1>-<\Psi_2,
\Phi_2>)+<\Lambda\Psi_2,
\Phi_1>+\phi_{N+1}^{2}\psi_{N+1}-\frac{1}{2}\phi_{N+1}\phi_{N+2}\psi_{N+2},\\
\phi_{N+2, t_2}=\frac{1}{4}\phi_{N+2}(<\Psi_1, \Phi_1>-<\Psi_2,
\Phi_2>)+\frac{1}{2}(<\Lambda\Psi_3, \Phi_1>-<\Lambda\Psi_2,
\Phi_3>),\\
\psi_{1j,
t_2}=(\lambda_j^{2}-\frac{1}{2}\phi_{N+1}\psi_{N+1}+\frac{1}{4}\phi_{N+2}\psi_{N+2})\psi_{1j}
-(\psi_{N+1}\lambda_j-\frac{1}{2}<\Psi_1,
\Phi_2>)\psi_{2j}\\\qquad\quad-\frac{1}{4}(2\psi_{N+2}\lambda_j+
<\Psi_1, \Phi_3>+<\Psi_3, \Phi_2>)\psi_{3j},\\
\psi_{2j, t_2}=-(\phi_{N+1}\lambda_j-\frac{1}{2}<\Psi_2,
\Phi_1>)\psi_{1j}
+(-\lambda_j^{2}+\frac{1}{2}\phi_{N+1}\psi_{N+1}-\frac{1}{4}\phi_{N+2}\psi_{N+2})\psi_{2j}\\\qquad\quad
-\frac{1}{4}(2\phi_{N+2}\lambda_j+<\Psi_2, \Phi_3>-<\Psi_3, \Phi_1>)\psi_{3j},\\
\psi_{3j, t_2}=-\frac{1}{4}(2\phi_{N+2}\lambda_j+<\Psi_2,
\Phi_3>-<\Psi_3, \Phi_1>)\psi_{1j}+\frac{1}{4}(2\psi_{N+2}\lambda_j+
<\Psi_1, \Phi_3>+<\Psi_3,
\Phi_2>)\psi_{2j},\\
\psi_{N+1, t_2}=-<\Lambda\Psi_1,
\Phi_2>-\frac{1}{2}\psi_{N+1}(<\Psi_1, \Phi_1>-<\Psi_2,
\Phi_2>)-\phi_{N+1}\psi_{N+1}^{2}+\frac{1}{2}\phi_{N+2}\psi_{N+1}\psi_{N+2},\\
\psi_{N+2, t_2}=\frac{1}{2}(<\Lambda\Psi_1, \Phi_3>+<\Lambda\Psi_3,
\Phi_2>)-\frac{1}{4}\psi_{N+2}(<\Psi_1, \Phi_1>-<\Psi_2,
\Phi_2>)-\frac{1}{2}\phi_{N+1}\psi_{N+1}\psi_{N+2},\end{array}\right.\end{equation}
where $1\leq j\leq N$.

It is a direct but tedious check that  the nonlinearized super
system (\ref{21}) can be written as the following super Hamiltonian
form
\begin{equation}\left\{\begin{array}{l} \Phi_{1,
t_2}=\frac{\partial H_2}{\partial\Psi_1}, \quad\Phi_{2,
t_2}=\frac{\partial H_2}{\partial\Psi_2}, \quad\Phi_{3,
t_2}=\frac{\partial H_2}{\partial\Psi_3}, \quad\phi_{N+1,
t_2}=\frac{\partial H_2}{\partial\psi_{N+1}},\quad
\phi_{N+2, t_2}=\frac{\partial H_2}{\partial\psi_{N+2}},\\
\Psi_{1, t_2}=-\frac{\partial H_2}{\partial\Phi_1}, \quad\Psi_{2,
t_2}=-\frac{\partial H_2}{\partial\Phi_2}, \quad\Psi_{3,
t_2}=\frac{\partial H_2}{\partial \Phi_3}, \quad\psi_{N+1,
t_2}=-\frac{\partial H_2}{\partial\phi_{N+1}}, \quad\psi_{N+2,
t_2}=\frac{\partial
H_2}{\partial\phi_{N+2}},\end{array}\right.\end{equation} where the
super Hamiltonian is given by
\begin{eqnarray*}
H_2&=&-<\Lambda^{2}\Psi_1, \Phi_1>+<\Lambda^{2}\Psi_2,
\Phi_2>+\phi_{N+1}<\Lambda\Psi_1, \Phi_2>+\psi_{N+1}<\Lambda\Psi_2,
\Phi_1>\\&&+\frac{1}{2}\phi_{N+2}(<\Lambda\Psi_1,
\Phi_3>+<\Lambda\Psi_3,
\Phi_2>)-\frac{1}{2}\psi_{N+2}(<\Lambda\Psi_2,
\Phi_3>-<\Lambda\Psi_3, \Phi_1>)\\&&+
\frac{1}{4}(2\phi_{N+1}\psi_{N+1}-\phi_{N+2}\psi_{N+2})(<\Psi_1,
\Phi_1>-<\Psi_2, \Phi_2>)-\frac{1}{2}<\Psi_2, \Phi_1><\Psi_1,
\Phi_2>\\&&+\frac{1}{4}(<\Psi_2, \Phi_3>-<\Psi_3, \Phi_1>)(<\Psi_1,
\Phi_3>+<\Psi_3,
\Phi_2>)-\frac{1}{2}\phi_{N+1}\phi_{N+2}\psi_{N+1}\psi_{N+2}\\&&+\frac{1}{2}\phi_{N+1}^{2}\psi_{N+1}^{2}.\end{eqnarray*}

That is to say, as for $t_2$-part, the nonlinearized super system
(\ref{21}) is finite-dimensional super Hamiltonian system. In what
follow, we want to prove that for any $n\geq2$, the super system
(\ref{34}) and (\ref{34*}) can be nonlinearized, and furthermore,
the obtained nonlinearized system is finite-dimensional super
Hamiltonian system. Therefore, making use of (\ref{14*}) and Eq.
(\ref{L1}), we obtain the constrained $a_i, b_i, c_i, \rho_i,
\delta_i (1\leq i\leq N)$ in systems (\ref{34}) and (\ref{34*}).
Only for differentiation, $\tilde{P}(U)$ denotes the new expression
generated from $P(U)$ by the nonlinear constraint (\ref{15}). i. e.
\begin{equation}\label{12}\left\{\begin{array}{l}
\tilde{a}_i=-\frac{1}{4}<\Lambda^{i-2}\Psi_1,
\Phi_1>-\frac{1}{2}<\Lambda^{i-2}\Psi_2, \Phi_2>,
\quad i\geq2,\\
\tilde{b}_i=-\frac{1}{2}<\Lambda^{i-2}\Psi_2, \Phi_1>,\quad i\geq2,\\
\tilde{c}_i=-\frac{1}{2}<\Lambda^{i-2}\Psi_1, \Phi_2>,\quad i\geq2,\\
\tilde{\rho}_i=\frac{1}{4}(<\Lambda^{i-2}\Psi_2,
\Phi_3>-<\Lambda^{i-2}\Psi_3, \Phi_1>),
\quad i\geq2,\\
\tilde{\delta}_i=-\frac{1}{4}(<\Lambda^{i-2}\Psi_1,
\Phi_3>+<\Lambda^{i-2}\Psi_3, \Phi_2>),\quad i\geq2.
\end{array}\right.\end{equation}

Substituting (\ref{12}) into the super systems (\ref{34}) and
(\ref{34*}), we obtain the nonlinearized super system
\begin{equation}\label{tn}\left\{\begin{array}{l}
\left(\begin{array}{c}
\phi_{1j}\\\phi_{2j}\\\phi_{3j}\end{array}\right)_{t_n}
=\left(\begin{array}{ccc}
\sum\limits_{i=0}^{n}\tilde{a}_i\lambda_j^{n-i}&\sum\limits_{i=0}^{n}\tilde{b}_i\lambda_j^{n-i}
&\sum\limits_{i=0}^{n}\tilde{\rho}_i\lambda_j^{n-i}\\\sum\limits_{i=0}^{n}\tilde{c}_i\lambda_j^{n-i}
&-\sum\limits_{i=0}^{n}\tilde{a}_i\lambda_j^{n-i}&\sum\limits_{i=0}^{n}\tilde{\delta}_i\lambda_j^{n-i}\\
\sum\limits_{i=0}^{n}\tilde{\delta}_i\lambda_j^{n-i}&-\sum\limits_{i=0}^{n}\tilde{\rho}_i\lambda_j^{n-i}&0
\end{array}\right)\left(\begin{array}{c}\phi_{1j}\\\phi_{2j}\\\phi_{3j}\end{array}\right)
,\quad 1\leq j\leq N,\\ \left(\begin{array}{c}
\psi_{1j}\\\psi_{2j}\\\psi_{3j}\end{array}\right)_{t_n}
=\left(\begin{array}{ccc}
-\sum\limits_{i=0}^{n}\tilde{a}_i\lambda_j^{n-i}&-\sum\limits_{i=0}^{n}\tilde{c}_i\lambda_j^{n-i}
&\sum\limits_{i=0}^{n}\tilde{\delta}_i\lambda_j^{n-i}\\-\sum\limits_{i=0}^{n}\tilde{b}_i\lambda_j^{n-i}
&\sum\limits_{i=0}^{n}\tilde{a}_i\lambda_j^{n-i}&-\sum\limits_{i=0}^{n}\tilde{\rho}_i\lambda_j^{n-i}\\
-\sum\limits_{i=0}^{n}\tilde{\rho}_i\lambda_j^{n-i}&-\sum\limits_{i=0}^{n}\tilde{\delta}_i\lambda_j^{n-i}&0
\end{array}\right)\left(\begin{array}{c}\psi_{1j}\\\psi_{2j}\\\psi_{3j}\end{array}\right)
,\quad 1\leq j\leq N.\end{array}\right.\end{equation} In what
follows, we want to see that the nonlinearized super system
(\ref{tn}) is a  finite-dimensional super Hamiltonian system.

 From Eq. (\ref{12}), we know that the constrained co-adjoint representation
equation $\tilde{N}_x=[\tilde{M}, \tilde{N}]$ is still satisfied,
and furthermore, the equality $(\tilde{N}^{2})_x= [\tilde{M},
\tilde{N}^{2}]$ is also satisfied. Therefore, let
$$\tilde{F}=\frac{1}{2}Str\tilde{N}^{2}=\tilde{a}^{2}+\tilde{b}\tilde{c}+2\tilde{\rho}\tilde{\delta}.$$
It is not difficult to calculate that $\tilde{F}_x=0,$ which means
that $\tilde{F}$ is a generating function of integrals of motion for
the nonlinearized spatial system (\ref{16}). Let
$\tilde{F}=\sum\limits_{n\geq0}\tilde{F}_n\lambda^{-n}$, integrals
of motion $\tilde{F}_n (n\geq0)$ is given by the following formulas
 \begin{eqnarray}
 \tilde{F}_0&=&1,\qquad\tilde{F}_1=0,\nonumber\\
\tilde{F}_2&=&\frac{1}{2}(<\Psi_1, \Phi_1>-<\Psi_2,
 \Phi_2>)+\phi_{N+1}\psi_{N+1}-\frac{1}{2}\phi_{N+2}\psi_{N+2},\nonumber\\
 \tilde{F}_3&=&\frac{1}{2}(<\Lambda\Psi_1, \Phi_1>-<\Lambda\Psi_2,
 \Phi_2>)-\frac{1}{2}\phi_{N+1}<\Psi_1, \Phi_2>-\frac{1}{2}\psi_{N+1}<\Psi_2,
 \Phi_1>\nonumber\\&&-\frac{1}{4}\phi_{N+2}(<\Psi_1, \Phi_3>+<\Psi_3,
 \Phi_2>)+\frac{1}{4}\psi_{N+2}(<\Psi_2, \Phi_3>-<\Psi_3,
 \Phi_1>),\nonumber\\
\tilde{F}_n&=&\sum\limits_{i=2}^{n-1}[\frac{1}{16}(<\Lambda^{i-2}\Psi_1,
\Phi_1>-<\Lambda^{i-2}\Psi_2, \Phi_2>)(<\Lambda^{n-i-2}\Psi_1,
\Phi_1>-<\Lambda^{n-i-2}\Psi_2,
\Phi_2>)\nonumber\\&&-\frac{1}{8}(<\Lambda^{i-2}\Psi_2,
\Phi_3>-<\Lambda^{i-2}\Psi_3, \Phi_1>)(<\Lambda^{n-i-2}\Psi_1,
\Phi_3>+<\Lambda^{n-i-2}\Psi_3,
\Phi_2>)\nonumber\\&&+\frac{1}{4}<\Lambda^{i-2}\Psi_2,
\Phi_1><\Lambda^{n-i-2}\Psi_1,
\Phi_2>]+\frac{1}{2}(<\Lambda^{n-2}\Psi_1,
\Phi_1>-<\Lambda^{n-2}\Psi_2,
\Phi_2>)\nonumber\\&&-\frac{1}{4}\phi_{N+2}(<\Lambda^{n-3}\Psi_1,
\Phi_3>+<\Lambda^{n-3}\Psi_3,
\Phi_2>)+\frac{1}{4}\psi_{N+2}(<\Lambda^{n-3}\Psi_2,
\Phi_3>-<\Lambda^{n-3}\Psi_3,
\Phi_1>)\nonumber\\&&-\frac{1}{2}\phi_{N+1}<\Lambda^{n-3}\Psi_1,
\Phi_2>-\frac{1}{2}\psi_{N+1}<\Lambda^{n-3}\Psi_2, \Phi_1>,\qquad
n\geq4.\end{eqnarray}

After a direct calculation, we have
\begin{equation}\left\{\begin{array}{l}
\Phi_{1, t_n}=-2\frac{\partial F_{n+2}}{\partial\Psi_1},\quad
\Phi_{2, t_n}=-2\frac{\partial F_{n+2}}{\partial\Psi_2},\quad
\Phi_{3, t_n}=-2\frac{\partial F_{n+2}}{\partial\Psi_3},\quad
\phi_{N+1, t_n}=-2\frac{\partial F_{N+2}}{\partial\psi_{N+1}},\quad
\phi_{N+2, t_n}=-2\frac{\partial F_{N+2}}{\partial\psi_{N+2}},\\
\Psi_{1, t_n}=2\frac{\partial F_{n+2}}{\partial\Phi_1},\quad
\Psi_{2, t_n}=2\frac{\partial F_{n+2}}{\partial\Phi_2},\quad
\Psi_{3, t_n}=-2\frac{\partial F_{n+2}}{\partial\Phi_3},\quad
\psi_{N+1, t_n}=2\frac{\partial F_{N+2}}{\partial\phi_{N+1}},\quad
\psi_{N+2, t_n}=-2\frac{\partial F_{N+2}}{\partial\phi_{N+2}},
\end{array}\right.\end{equation}
which means that the nonlinearized temporal system (\ref{tn}) is
super Hamiltonian system. In conclusion, for any $n (n\geq1)$, the
nonlinearized system (\ref{tn}) is a finite-dimensional, super
Hamiltonian system. In what follows, we only want to prove that the
nonlinearized system (\ref{tn})  is completely integrable in the
Liouville sense.

 To this aim, we choose the following Poisson bracket
\begin{eqnarray}\label{poisson}\{ F, G\}&=&\sum_{i=1}^{3}\sum_{j=1}^{N}
(\frac{\partial F}{\partial\phi_{ij}}\frac{\partial
G}{\partial\psi_{ij}}-(-1)^{p(\phi_{ij})p(\psi_{ij})}\frac{\partial
F}{\partial\psi_{ij}}\frac{\partial
G}{\partial\phi_{ij}})\nonumber\\&+&\sum_{j=1}^{2}(\frac{\partial
F}{\partial\phi_{N+j}}\frac{\partial
G}{\partial\psi_{N+j}}-(-1)^{p(\phi_{N+j})p(\psi_{N+j})}\frac{\partial
F}{\partial\psi_{N+j}}\frac{\partial
G}{\partial\phi_{N+j}}),\end{eqnarray} where $p(u)$ is a parity
function of $u$, namely, $p(u)=0$ if $u$ is an even variable and
$p(u)=1$ if $u$ is an odd variable. It is not difficult to see that
$\tilde{F}_n (n\geq0)$ are  also integrals of motion for Eq.
(\ref{tn}), i. e.
$$\{ \tilde{F}_{m+1}, \tilde{F}_{n+2}\}=-\frac{1}{2}\frac{\partial}{\partial t_n}\tilde{F}_{m+1}=0,\quad  m,
n\geq0,$$ which means that $\{\tilde{F}_{n}\}_{n\geq0}$ are in
involution in pair.

With the help of the result of nonlinearization
 for classical integrable system \cite{ma94,ma95,ma96}, it is natural for us to set
\begin{equation}\label{fk}
f_k=\psi_{1k}\phi_{1k}+\psi_{2k}\phi_{2k}+\psi_{3k}\phi_{3k},\quad
1\leq k\leq N,
\end{equation}
 and  verify that they are also integrals of motion of the
constrained spatial system (\ref{16}) and temporal system
(\ref{tn}). Making use of (\ref{poisson}), it is easy to find that
(\ref{fk}) are in involution in pair. For the nonlinearized spatial
system (\ref{16}) and the nonlinearized temporal system (\ref{tn}),
we choose 3N+2 integrals of motion
\begin{equation}\label{3N+2}
f_1, \cdots, f_N, F_2, F_3, \cdots, F_{2N+3}, \end{equation} whose
involution have been verified. In what follows, we want to show the
functional independence of (\ref{3N+2}). Similar to the Refs.
\cite{Heyu,ma96,ma01}, 3N+2 functions (\ref{3N+2}) are functionally
independent at least over some region of the supersymmetry manifold
$\mathbb{R}^{4N+2|2N+2}$.

Taking into account  the preceding program, it is not difficult to
draw a below.
\begin{Theorem} The constrained (6N+4)-dimensioanl systems (\ref{16}) and (\ref{tn}) are  super Hamiltonian systems,
whose 3N+2 integrals of motion (\ref{3N+2})
 are in involution in pair and functionally independent over supersymmetry manifold $\mathbb{R}^{4N+2|2N+2}$.
\end{Theorem}
\begin{Remark}The main differences between the finite-dimensional super Hamiltonian systems in
present paper and reference \cite{Heyu} can be summarized bellow:

1) Due to the implicit constraint,  we have to introduce other 4
coordinates in Eq.(\ref{new viables}) such that finite-dimensional
super Hamiltonian system  in present paper is (6N+4)-dimensional.
However, the corresponding system in reference \cite{Heyu} is
6N-dimensional.

2)Correspondingly, Poisson bracket in Eq.(\ref{poisson}) is
different from Eq.(36) in Ref.\cite{Heyu}.
\end{Remark}

\section{Conclusions and Discussions}

In this paper, we presented a new finite-dimensional integrable
super Hamiltonian system of the super AKNS system. The difference of
this paper and Ref. \cite{Heyu} lies on the symmetry constraint
between the potentials and the eigenfunctions, which results in
generating different finite-dimensional super system. In Ref.
\cite{Heyu}, we have proposed an explicit symmetry constraint.
Substituting the explicit constraint into the spatial system and the
temporal system of the super AKNS system, we have obtained the
nonlinearized super system, and furthermore, we have proved the
obtained nonlinearized system are complete integrable in the
Liouville sense. However in this paper, we proposed an implicit
symmetry constraint (\ref{15}), which made the potentials can not be
expressed by eigenfunctions explicitly. Therefore, we refer to the
method of implicit constraint for classical integrable system, and
introduce four new variables (\ref{new viables}) to explicitly
express potentials. After this, we obtained the super nonlinearized
spatial system (\ref{16}) and the temporal system (\ref{tn}), and
proved that the obtained super system (\ref{16}) and (\ref{tn}) is
super Hamiltonian system and have 3N+2 integrals of motion, which is
in involution in pair and functionally independent at least over
some region of supersymmetry manifold $\mathbb{R}^{4N+2|2N+2}$. At
last, we would like to stress that the nonlinearization  of the
super AKNS system provides a new and systematic way to construct
finite-dimensional super Hamiltonian system, and there are very few
examples \cite{landi} of Hamiltonian system with fermionic variables
in literatures. Additionally, to illustrate the potential
applications of the finite-dimensional super Hamiltonian system
obtained in previous sections, we would like to show following two
points of physics and mathematics, respectively. On the one hand, it
is possible to find these systems in the finite-dimensional super
physical theory in the future. For example, the super analogue of
the integrable Rosochatius deformation because finite-dimensional
integrable Rosochatus systems, which are important integrable
structures in string theory, can be obtained \cite{zhou2007} through
nonlinearization of the AKNS system. On the other hand,
finite-dimensional C. Neumann system \cite{cao1995,zhou1998}, which
describes the motion of a particle on $S^{N-1}$ with a quadratic
potential in $N$-dimensional space, is derived again from the AKNS
by nonlinearization. So, it is possible to establish corresponding
super C. Neumann in the supersymmetry case through the
nonlinearization of the super AKNS system.

However,  for the super AKNS system, we have not yet found another
kind of implicit symmetry constraint, which will engender
finite-dimensional integrable super Hamiltonian system. The
difficulty lies in the selection of new variables. Once we find
proper new variables, the method of nonlinearization for the super
AKNS system under new implicit symmetry constraint will be carried
out. In addition, under implicit constraints, nonlinearization of
the other super systems   will be studied in our future work.

{\bf Acknowledgments} {\small This work is supported by the Hangdian
Foundation KYS075608072 and KYS075608077, NSF of China under grant
number 10681187. He is also supported by the Program for NCET under
Grant No.NCET-08-0515. We thank anonymous referees for their
valuable suggestions and pertinent criticisms. }


\small \baselineskip 13pt
\begin{thebibliography}{99} %\cite{ZS}
%%nonlinearization
\bibitem{cao88} C. W. Cao, Nonlinearization of the Lax system for AKNS hierarchy, Sci. China Ser. A 33 (1990),
no. 5, 528-536.
\bibitem{Konopelchenko} B. Konopelchenko, J. Sidorenko, W. Strampp,
(1+1)-dimensional integrable systems as symmetry constraints of
(2+1)-dimensional systems,  Phys. Lett. A 157 (1991), no. 1, 17-21.
\bibitem{cheng91} Y. Cheng, Y. S. Li, The constraint of the Kadomtsev-Petviashvili equation and its special solutions,
Phys. Lett. A 157 (1991), no. 1, 22-26.
\bibitem{cheng92} Y. Cheng, Constraints of the Kadomtsev-Petviashvili hierarchy, J. Math. Phys. 33 (1992), no. 11, 3774-3782.
\bibitem{ma94} W. X. Ma, W. Strampp, An explicit symmetry constraint for the Lax pairs and the adjoint Lax pairs of
AKNS systems, Phys. Lett. A 185 (1994), no. 3, 277-286.
\bibitem{zeng95} Y. B. Zeng, The integrable system associated with higher-order constraint, Acta Mathematica Sinica, 38 (1995), no. 5, 642-652.
\bibitem{ma00} Y. S. Li, W. X. Ma, Binary nonlinearization of AKNS spectral problem under higher-order symmetry
constraints, Chaos, Solitons and Fractals 11 (2000), no. 5, 697-710.
\bibitem{yu08} Y. You, J. Yu, Q. Y. Jiang, An implicit symmetry constraint of the modified Korteweg-de Vries
(mKdV) equation, J. Zhejiang University Science A,  9 (2008), no.
10, 1457-1462.
\bibitem{Gurses85} M. Gurses, O. Oguz, A super AKNS scheme, Phys. Lett. A 108 (1985), no. 9, 437-440.
\bibitem{Li86} Y. S. Li, L. N. Zhang, Super AKNS scheme and its infinite conserved currents, Nuovo Cimento A 93 (1986), no. 2, 175-183;
A note on the super AKNS equations, J. Phys. A 21 (1988), no. 7,
1549-1552.
\bibitem{Shaw98} J. C. Shaw, M. H. Tu, Canonical gauge equivalences of the sAKNS and sTB hierarchies, J. Phys. A 31 (1998), no. 30, 6517-6523.
\bibitem{popowicz} Z. Popowicz, The fully supersymmetric AKNS equations, J. Phys. A 23 (1990), no. 7, 1127-1136.
\bibitem{aratyn99} H. Aratyn, E. Nissimov, S. Pacheva, Supersymmetric Kadomtsev-Petviashvili hierarchy: "ghost" symmetry
structure, reductions, and Darboux-B$\ddot{a}$cklund solutions, J.
Math. Phys. 40 (1999), no. 6, 2922-2932.
\bibitem{ghost03} S. Ghosh, D. Sarma, Soliton solutions of the N=2 supersymmetric KP equation, J. Nonlinear
 Math. Phys. 10 (2003), no. 4, 526-538.
\bibitem{liu97} Q. P. Liu, M. Man$\tilde{a}$s, Darboux transformation for the Manin-Radul supersymmetric KdV equation,
Phys. Lett. B 394 (1997), 337-342.
\bibitem{liu00} Q. P. Liu, M. Man$\tilde{a}$s, Darboux transformations for super-symmetric KP hierarchies, Phys. Lett. B 485 (2000), no. 1-3, 293-300.
\bibitem{li90} Y. S. Li, L. N. Zhang, Hamiltonian structure of the super evolution equation, J. Math. Phys. 31 (1990), no. 2, 470-475.
\bibitem{oevel91} W. Oevel, Z. Popowicz, The bi-Hamiltonian structure of fully supersymmetric Korteweg-de Vries systems,
Commun. Math. Phys. 139 (1991), 441-460.
\bibitem{tu99} M. H. Tu, J. C. Shaw, Hamiltonian structures of generalized Manin-Radul super-KdV and constrained super KP hierarchies,
J. Math. Phys. 40 (1999), no. 6, 3021-3034.
\bibitem{Heyu} J. S. He, J. Yu, Y. Cheng, R. G. Zhou, Binary nonlinearization of the super AKNS system,
Modern Phys. Lett. B 22 (2008), no. 4, 275-288.
\bibitem{cartier}P. Cartier, C. DeWitt-Morette, M. Ihl and C. S$\ddot{a}$mann,
Supermanifolds-application to supersymmetry, in {\sl Maultiple Facts of Quantization and Supersymmetry}, edited by
M. Olshanetsky and A. Vanishetin( World Sci. Publ., River Edge, NJ, 2002), 412-457.

\bibitem{ma95} W. X. Ma,  Symmetry constraint of MKdV equations by binary nonlinearization, Physica A 219 (1995), 467-481.
\bibitem{ma96} W. X. Ma, B. Fuchssteiner, W. Oevel, A 3$\times$3 matrix spectral problem for AKNS hierarchy and
its binary nonlinearization, Phys. A 233 (1996), 331-354.
\bibitem{ma01} W. X. Ma, Z. X. Zhou, Binary symmetry constraints of N-wave interaction equations in 1+1 and
2+1 dimensions, J. Math. Phys. 42 (2001), no. 9, 4345-4382.
\bibitem{landi}G. Landi, G. Marmo, G. Vilasi, Remarks on the complete integrability of the dyanmical systems
with the fermionic variables, J. Phys. A 25 (1992), 4413-4423.
\bibitem{zhou2007} R. G. Zhou, Integrable Rosochatisu deformations of the rsetricted soltion flows,
J. Math. Phys 48 (2007), 103510, 17pages.
\bibitem{cao1995} C. W. Cao, Classical Integrable Systems, in {\sl Soliton Theory and its Applications} Edited by
C. H. Gu (Springer-Verlag(Berlin, Germany)and Zhejiang Science and
Technology Publising House(Hangzhou, China), 1995) p.152-191.
\bibitem{zhou1998} R. G. Zhou, r-Matrix for the Restricted KdV Flows with the Neumann Constraints, J. Nonlinear Math.
Phys. 5 (1998), 181-189.




%% Remarks on the complete integrability of dynamical systems with ferimonic variables.
\end{thebibliography}
\end{document}